\documentclass[10.75pt, a4paper]{article}

\usepackage[utf8]{inputenc}
\usepackage{geometry}
\usepackage[citestyle=numeric,style=phys,biblabel=brackets,backend=bibtex,sorting=none,url=false,doi=false, natbib]{biblatex}

\usepackage[noend]{algpseudocode}
\usepackage{algorithm}
\usepackage{amsmath}
\usepackage{amssymb}
\usepackage{nicefrac}
\usepackage{siunitx}
\usepackage{bm}
\usepackage[tbtags]{mathtools}

\usepackage{abstract}
\usepackage{multirow}
\usepackage{tabularx}
\usepackage{booktabs}
\usepackage{array}
\usepackage{longtable}
\usepackage[affil-it]{authblk} 
\usepackage[]{parskip} 
\usepackage{dcolumn}
\usepackage{lipsum}
\usepackage{titlesec}

\usepackage[font=footnotesize]{caption}

\usepackage{graphicx}
\usepackage[usenames,dvipsnames]{xcolor}
\usepackage{subfigure}
\usepackage{color, colortbl}
\usepackage{wrapfig}
\usepackage{float}

\usepackage{etoolbox}
\usepackage{eurosym}
\usepackage[hidelinks]{hyperref}
\usepackage{csquotes}
\usepackage{etoolbox}
\usepackage{lmodern}
\usepackage{enumerate}
\usepackage{enumitem}
\usepackage[normalem]{ulem}
\usepackage{csvsimple}
\usepackage{changepage}
\usepackage{csquotes}
\usepackage[
	nonumberlist, 				
	acronym,      				
	nomain,						
	nopostdot,					
]{glossaries}
\usepackage{glossary-superragged}
\usepackage{ellipsis}
\usepackage[framemethod=tikz]{mdframed}
\usepackage{dirtytalk}
\usepackage{tcolorbox}
\usepackage{url}
\usepackage{pifont}
\usepackage{soul}

\titleformat{\section}
  {\bf\sffamily}
  {\thesection. }
  {5pt}
  {\MakeUppercase}
\renewcommand{\thesection}{\Roman{section}} 

\titleformat{\subsection}
  {\bf\sffamily}
  {\thesubsection. }
  {5pt}{}
  
\renewcommand{\thesubsection}{\Alph{subsection}}

\DefineBibliographyStrings{english}{andothers={\itshape et\addabbrvspace al\adddot}}

\makeatletter
\patchcmd{\@maketitle}{\LARGE \@title}{\fontsize{16}{19.2}\selectfont\@title}{}{}
\makeatother

\let\vec\mathbf

\newcolumntype{L}[1]{>{\raggedright\let\newline\\\arraybackslash\hspace{0pt}}m{#1}}
\newcolumntype{C}[1]{>{\centering\let\newline\\\arraybackslash\hspace{0pt}}m{#1}}
\newcolumntype{R}[1]{>{\raggedleft\let\newline\\\arraybackslash\hspace{0pt}}m{#1}}

\newtcolorbox{mybox}[1]{colback=green!6!white,colframe=black!75!black,fonttitle=\bfseries,title=#1}
\newtcolorbox{mybox2}{colback=red!5!white,colframe=red!75!black}

\setstcolor{red}

\hypersetup{
   colorlinks,
   linkcolor={blue!50!black},
   citecolor={blue!50!black},
   urlcolor={blue!80!black}
} 

\renewcommand{\vec}[1]{{\bm #1}}
\newcommand{\mat}[1]{\bm{\mathsf{#1}}}

\newcommand{\pderiv}[2]{\dfrac{\partial#1}{\partial#2}}
\newcommand{\ptderiv}[1]{\pderiv{#1}{t}}

\definecolor{brickred}{rgb}{0.8, 0.25, 0.33}
\definecolor{darkorange}{rgb}{1.0, 0.55, 0.0}
\definecolor{persiangreen}{rgb}{0.0, 0.65, 0.58}
\definecolor{persianindigo}{rgb}{0.2, 0.07, 0.48}
\definecolor{cadet}{rgb}{0.33, 0.41, 0.47}
\definecolor{turquoisegreen}{rgb}{0.63, 0.84, 0.71}
\definecolor{sandybrown}{rgb}{0.96, 0.64, 0.38}
\definecolor{blueblue}{rgb}{0.0, 0.2, 0.6}
\definecolor{ballblue}{rgb}{0.13, 0.67, 0.8}
\definecolor{greengreen}{rgb}{0.0, 0.5, 0.0}
\definecolor{mycolor}{rgb}{0.122, 0.435, 0.698}

\bibliography{references}

\title{Rheology of Parametric Odd Materials}

\usepackage{authblk}
\usepackage{orcidlink}

\author[1]{Pedro Cosme\thanks{ p.a.cosmeesilva@uva.nl; ORCID: \href{https://orcid.org/0000-0001-6754-7762}{0000-0001-6754-7762}}}
\author[1,2]{Maziyar Jalaal\thanks{m.jalaal@uva.nl; m.jalaal@damtp.cam.ac.uk; ORCID: \href{https://orcid.org/0000-0002-5654-8505}{0000-0002-5654-8505}}}

\affil[1]{Van der Waals-Zeeman Institute, Institute of Physics, University of Amsterdam,\protect\\
Science Park 904, 1098 XH Amsterdam, The Netherlands}
\affil[2]{Department of Applied Mathematics and Theoretical Physics, University of Cambridge,\protect\\
Wilberforce Road, Cambridge CB3 0WA, United Kingdom}

\begin{document}
\begingroup
\sffamily

\date{}

\maketitle
\endgroup

\begin{abstract}
The design and control of rheological properties in active systems are crucial for developing functional materials with novel properties. While significant attention has been devoted to odd viscosity and odd elasticity, comprehensive odd rheological models remain scarce. Here, we introduce and characterise a platform-agnostic odd viscoelastic material, where complex responses arise from parametric control of internal rotational degrees of freedom and are triggered by coupling between normal and shear stresses. Such control can be achieved through either parametric driving or a parametric nonlinear feedback. We further show that this coupling can induce rotation-driven fluidisation, transitioning the material from an elasticity-dominated to a viscosity-dominated regime. Lastly, we show that the interplay between odd and even sectors enhances the nonlinear material response under large-amplitude oscillatory shear (LAOS). The findings outline a route to Floquet-engineered metafluids with programmable, non-reciprocal rheology.
\end{abstract}

\maketitle

\section{Introduction}

Beyond Newtonian rheology, there is a large realm of materials that exhibit a complex response to applied stimuli. Within it, one encounters systems where the boundaries of solid and liquid behaviour fade, and the materials can present mechanical changes in response to the amount of strain or strain-rate or to the history of the perturbations.
This dual character defines a  large class of 
materials, encompassing systems such as colloidal gels, emulsions, foams, pastes, and many soft glassy systems~\cite{balmforth2014yielding,bonn2017yield}, where, in general, the complex rheology is determined by complex microscopic configurations or dynamics. At the continuum level, such materials are described by constitutive laws in which stress couples symmetrically to strain and strain rate, consistent with the parity and time-reversal symmetries~\cite{LandauLifshitzFluid}.
By contrast, materials that break parity or time-reversal symmetry exhibit antisymmetric components in their transport tensors~\cite{avron_odd_1998,Banerjee2017OddViscosity,Fruchart2023Review}. In these systems, often featuring odd viscosity and odd elasticity, stress can develop components transverse to strain rate or strain, enabling phenomena such as nondissipative energy transfer, Hall-like responses, and unconventional wave propagation~\cite{avron_odd_1998, Fruchart2023Review, Soni2019TheFluid,Tan2022OddCrystals,duPlessis2026}. 
%
Beyond purely viscous or purely elastic responses, odd viscoelasticity, where elastic and viscous sectors are coupled, has recently attracted significant attention~\cite{Banerjee2021,lier2022passive,floyd2024pattern,BanarjeeSolich2025}. In these descriptions, the odd coupling enters as a fixed material constant, and how the rheological state of an odd material, its solid- or fluid-like behaviour, can be controlled, switched or made to respond to stress through the chiral activity itself remains unexplored. 
This is particularly interesting, as recent advances in odd elastic solids with topological defects~\cite{Braverman2021OddElasticity} and in odd granular systems with plastic rearrangements~\cite{huang2023odd}, together with experimental platforms such as externally driven chiral colloids~\cite{bililign2022motile,guillet2025melting}, chiral active biological \say{particles}~\cite{Tan2022OddCrystals,eisenmann2026chirality}, and programmable robotic matter~\cite{saintyves2024self,veenstra2025adaptive,Zheng2025MetamaterialRobotics}, highlight the potential of parity-violating mechanics for tunable functionality, including those with viscoelastic properties.

Here, we formulate a minimal continuum theory of odd viscoelasticity in which the coupling between shear and normal-difference stresses arises from the microscopic rotations, and can therefore turned on and off or modulated in a functional form. When such rotation is constant, the model is the familiar odd Maxwell fluid of gyroviscous and magnetised systems~\cite{Kaufman1960PlasmaField,bernsteinOdd} and the linear limit of the recent odd viscoelastic models~\cite{Banerjee2021,lier2022passive,floyd2024pattern,BanarjeeSolich2025}. We let the coupling vary in time and follow the instantaneous moduli as the material is driven through its parametric response. Furthermore, we let it switch on under stress and find that the material fluidises on its own, within a finite window of driving frequencies; and we look at the nonlinear large-amplitude response that such a switch produces. To this end, we bring the tools of linear and nonlinear rheology to odd materials.

As a starting point, we consider an abstract fluid material in order to study its viscous fluid to elastic solid transition, at all sectors of the response modulus. At the linear response level, the stress tensor of a flow can be written as $\sigma_{ij}={\sigma}_{ij}^{(0)}+{ C_{ijkl}\, e_{kl} +\eta_{ijkl}\, \dot e_{kl}}$, resorting to rank-four viscosity and elasticity tensors ($\bm \eta$ and $\bm C$, respectively) acting on the strain $\vec{e}$ and strain rate $\dot{\vec{e}}$ and where $\sigma^{(0)}$ encodes hydrostatic contributions such as pressure and torques. Although for conventional fluids, such a tensor is symmetric over $ij\leftrightarrow kl$, under time reversal and parity breaking, it can display antisymmetric blocks that are overall termed \emph{odd}, viscosities, elasticities, or in general moduli. As the parity breaking implies, this requires a clear axial direction and while it can exist in three dimensions (3D)  with the appropriate symmetry constraints~\cite{avron_odd_1998,Khain2022StokesViscosities}, we focus on two-dimensional (2D) flows where the main implications can already be observed and the implementation is more straightforward. 
The stress tensor of a 2D isotropic fluid, in its most general form, can then be characterised by six viscosity parameters, and this number gets reduced to four in incompressible scenarios. However, the terms that contribute to the antisymmetric part of the stress tensor equilibrate on a typically faster timescale to obey entropy production rate~\cite{Curtiss1981ClassicalSections,Mccourt1965TransportII,Curtiss1956KineticMolecules,Kagan1962ONFREEDOM}. For that reason, the most relevant terms are those that contribute to a symmetric stress tensor: the well-known shear viscosity and the odd viscosity proper.  
Likewise, the elasticity tensor $C_{ijkl}$ also follows the same representation decomposition. As such, it would be tempting to think that in a viscoelastic setting, each sector could be described independently: shear strain governed by a Maxwell model with certain parameters, while the odd or antisymmetric parts would follow distinct governing models with separate parameters. However, as we will show, that is not the case, and each part influences the others. Consequently, one may, in principle, control or engineer the odd properties of a material to modify or act on its shear moduli.

\section{Odd Maxwell fluids: A Coupled odd rheological model}

\begin{figure}[!ht]
    \centering
    \includegraphics[width=\linewidth]{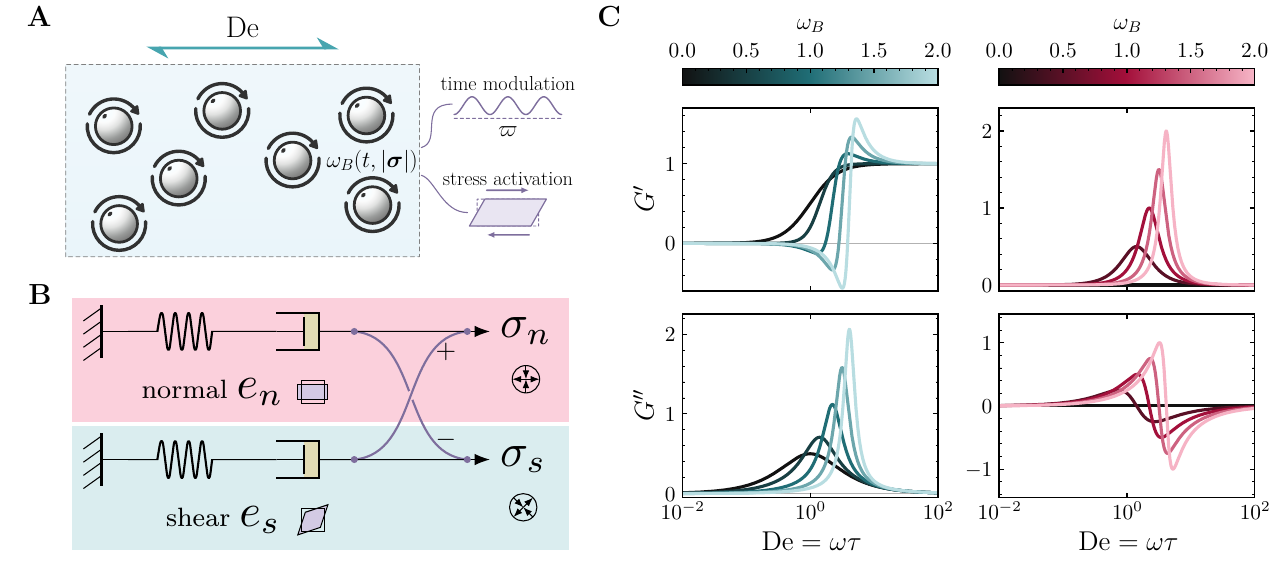}
        \caption{\textbf{A.} Fluid of chiral particles of the same handedness under oscillatory shear of dimensionless frequency $\mathrm{De}$. The effective rotational coupling $\omega_B(t,|\vec{\sigma}|)$ can be controlled through time modulation at frequency $\varpi$ or stress activation. \textbf{B.} Linked spring-dashpot representation of the normal (pink) and shear (teal) sectors, with their cross-coupling represented by the connections labelled $+$ and $-$. \textbf{C.} Storage moduli $G^{\prime}$ (top) and loss moduli $G^{\prime\prime}$ (bottom) in the shear (left) and normal (right) sectors as a function of the Deborah number for different rotational couplings $\omega_B$; the black curves representing the classical Maxwell model recovered in the limit $\omega_B\to0$.}
    \label{fig:intro}
\end{figure}

As a minimal picture, we envision a material made from a collection of chiral constituents and considerably free, so that it rests at a fluidised state. The internal rotational dynamics provide an additional degree of freedom for controlling the mechanical response (Fig. \ref{fig:intro}A). Such control may arise either from a time modulation (that in turn can be externally imposed or derive from internal dynamics of the particle, like circadian rhythms in microorganisms) or from an intrinsic dependence of the microscopic rotation on the stress of the material. At the macroscopic level, this translates into a coupling between the normal and shear sectors, which can be represented schematically by two linked Maxwell elements (Fig.  \ref{fig:intro}B). This coupling allows the odd properties of the material to act directly on its conventional viscoelastic response.

%

{
We want to retrieve the evolution of the stress-pressure tensor $\Pi$ defined as the covariance of the microscopic velocities, i.e. 
$\Pi_{ij}=\langle c_ic_j \rangle =\int c_i c_j f\,{\rm d}\vec{v}=p\delta_{ij}-\sigma_{ij}$
with $f$ the distribution function of the ensemble and where $\vec{c}$ is the peculiar velocity $\vec{c}=\vec{v}-\langle\vec{v}\rangle=\vec{v}-\vec{u}$, and with the usual decomposition into its hydrostatic pressure and stress parts.
From the Boltzmann equation 
\begin{equation}
\ptderiv{f}+\vec{v}\cdot\bm{\nabla}_\vec{x}f+\vec{a}\cdot\bm{\nabla}_\vec{v}f=\mathfrak{S},
\end{equation}
in which we assume that the acceleration $\vec{a}$ includes both physical and fictitious contributions; multiplying by $c_ic_j$ and integrating over the microscopic velocities leads to the transport law
\begin{equation}
\frac{D\sigma_{ij}}{Dt}+ \sigma_{ij}\partial_ku_k+ \sigma_{ik}\partial_ku_j+ \sigma_{jk}\partial_ku_i-\sigma_{kl}\partial_lu_k\delta_{ij}= -\Omega_{ijkl}\sigma_{kl}-\frac{\sigma_{ij}}{\tau} +p\big(\partial_iu_j+\partial_ju_i-  \delta_{ij}\partial_ku_k  \big),
\label{eq:evolveSigma}
\end{equation}
where we used  Grad moments expansion~\cite{Grad1949,kremer_introduction_2010} to obtain the closure $\langle c_ic_jc_k\rangle=0$ and under BGK approximation $\mathfrak{S}=(f^{(\text{equilibrium})}-f)/\tau$~\cite{Chapman1954TheGases,Bhatnagar1954,Gross1956} which introduces the relaxation time. In fact, it can be shown that, once retaining fictitious contributions to $\vec{a}$, \eqref{eq:evolveSigma} is frame invariant, and that the LHS can be regarded as an objective time derivative obtained from kinetic description~\cite{Wagner2001,Struchtrup2003,Ottinger2020} (the non-objectivity sometimes reported stesm from the omission of the inertial accelerations \emph{cf.} Appendix~\ref{APP:objective}).  

There, the coupling tensor $\vec{\Omega}$ will be specific for each interaction and encodes the averaging from the acceleration terms
\begin{equation}
    \langle c_ia_j+c_ja_i\rangle+\langle c_ic_j \partial_{v_k} a_k\rangle\equiv\Omega_{ijkl}\sigma_{kl}.
\end{equation}

}
While in materials without an odd response $\vec{\Omega}$ is either absent or diagonal, here we are interested in cases where it is antisymmetric on the $ij\leftrightarrow
 kl$ commutation of indices, so that the viscosity tensor inherits such structure. In fact a nonsymmetric tensor $\vec{\Omega}$ results either from the integration of chiral force terms in the Boltzmann equation or from parity-violating collision terms. The former arises when the force acting on the fluid constituents depend on the velocity, at least linearly, as in the case of charged particles under a Lorentz force in an external magnetic field~\cite{Pongsangangan2024ShearGraphene,CosmeeSilva2025,Berdyugin2019}, or active particles with an intrinsic rotation suspended in a viscous medium (see Fig~\ref{fig:intro}A). The latter mechanism has been recently proposed for odd fluids~\cite{Fruchart2022TheTheoryMisc}. In the relaxation-time limit, it corresponds to a modified collision term of the form $\Lambda_{ijkl}\sigma_{kl}/\tau$, which can absorb into the $\Omega_{ijkl}$ term.
While such microscopic realisations, with a marked chirality, may seem hard to attain, especially in passive systems, they are now increasingly common in active matter systems ~\cite{Soni2019TheFluid,Zhang2020,Lou2022}, rough granular suspensions, where collisions often break parity~\cite{Han2021,Zhao2022-ht}, or in ensembles of chiral biological agents~\cite{Tan2022OddCrystals,eisenmann2026chirality}; and one can further envision synthetic metamaterials designed to have the appropriate response. 

Assuming thus that we are in the presence of such a chiral ensemble, from~\eqref{eq:evolveSigma}, to linear order in strain rate amplitude, can be cast into the a dimensional form
 \begin{equation}
\ptderiv{\sigma_{ij}}+\sigma_{ij}+\tau\, \Omega_{ijkl}\sigma_{kl}= {\rm Wi}\,\dot{ e}_{ij}, \label{eq:generalbgkvisco}
\end{equation}
with $\dot{ e}_{ij}\equiv \partial_iu_j+\partial_ju_i-  \delta_{ij}\partial_ku_k $ and where we set the time scale with the relaxation time $\tau$ and define the Weissenberg number
\begin{equation}
    {\rm Wi} = \frac{\tau p \dot e_0}{\sigma_0}, 
\end{equation} that encodes the typical amplitude of the shear rate. This constitutes a generalised Maxwell model for the viscous tensor, and as will become evident below, the allowed symmetries of $\vec{\Omega}$ play a crucial role.

Since an isotropic odd viscoelastic response is only allowed in two dimensions, we henceforth restrict the discussion to such dimensionality. Thus, we resort to a more suitable basis, common in the literature, for the second-rank tensors, spanned by  
\begin{equation}
\varsigma^b=\begin{pmatrix}
1&0\\0&1\end{pmatrix}
\quad\varsigma^r=\begin{pmatrix}
0&-1\\1&0\end{pmatrix}
\quad\varsigma^n=\begin{pmatrix}
1&0\\0&-1        \end{pmatrix}
\quad\varsigma^s=\begin{pmatrix}0&1\\1&0
    \end{pmatrix}.    
\end{equation}
On this representation, the general isotropic viscosity tensor follows the structure 
\begin{multline}
    \eta=\zeta\varsigma^b\otimes\varsigma^b+ \eta_A\varsigma^r\otimes\varsigma^b+ \eta_B\varsigma^b\otimes\varsigma^r+\eta_R\varsigma^r\otimes\varsigma^r+ \eta \big(\varsigma^n\otimes\varsigma^n+\varsigma^s\otimes\varsigma^s\big)+\eta_o\big(\varsigma^n\otimes\varsigma^s-\varsigma^s\otimes\varsigma^n\big)
\end{multline}
where $\zeta$, $\eta_R$ and $\eta$ are the classically well-known bulk, rotational, and shear viscosities, respectively. The remaining coefficients couple deformations of different natures, \emph{e.g.} rotational with compressive, with our focus on the odd viscosity proper, the $\eta_o$. Likewise, in the context of this work, any other rank-four tensor, as $\Omega_{ijlk}$ or $\Lambda_{ijkl}$, would have to abide by a similar structure to represent physical objects. Moreover, from~\eqref{eq:generalbgkvisco}, one can easily see that for the viscous tensor to be antisymmetric, the tensor $\bm \Omega$ needs to exhibit the desired $\varsigma^n\otimes\varsigma^s-\varsigma^s\otimes\varsigma^n$  
 component. We thus proceed with this assumption and see its consequences in coupling both shears and overall triggering unconventional rheological responses.

The direct consequence of the structure of this viscoelastic model is that the different sectors become coupled (see Fig~\ref{fig:intro}B). As stated, we assume the simplest antisymmetric form for $\vec{\Omega}$, writing it as 
\begin{equation}
\tau\,\Omega_{ijkl}=\omega_B\bigg(\varsigma_{ij}^s\varsigma_{kl}^n-\varsigma_{ij}^n\varsigma_{kl}^s\bigg),
\end{equation}
reminiscent of the case where the fluid constituents are under a constant Lorentz or Coriolis force. Writing it in terms of a dimensionless rotational coupling $\omega_B$, where the microscopic rotation frequency has been rescaled by the relaxation time $\tau$. Under this assumption, the shear and normal stresses evolve as 
\begin{subequations}
    \begin{gather}
    \dot \sigma_n+\sigma_n={\rm Wi}\, \dot e_n+2\omega_B \sigma_s,\\
    \dot \sigma_s+\sigma_s={\rm Wi}\, \dot e_s-2\omega_B \sigma_n,
\end{gather}\label{eq:coupledMaxwell}%
\end{subequations}%
where $ \sigma_n=\sigma_{xx}-\sigma_{yy}\quad \text{and} \quad \sigma_s=\sigma_{xy}+\sigma_{yx}$ and likewise for $\dot e_s$ and $\dot e_n$.
Thus, it is evident that shear and normal stresses behave as in viscoelastic fluids, coupled through the microscopical parity breaking. This system of first-order differential equations can be diagonalised into a second order one, bringing it closer to some recently proposed models of odd viscoelasticity of constant $\omega_B$, based on symmetry or single-particle kinematics arguments~\cite{bernsteinOdd,Kaufman1960PlasmaField,Banerjee2021,BanarjeeSolich2025}. In this constant-$\omega_B$ form, the stress oscillates at $O(\omega_B)$ during the memory time, and the model is equivalent to the linear odd Maxwell models of those works; in contrast, what follows exploits the kinetic origin of the coupling, which allows it to be made time- and stress-dependent, leading to parametrically controlled rheology and to autonomous rotation-induced fluidisation.
In Fourier space the linear response
\begin{equation}
    \begin{pmatrix}
        \sigma_n\\
        \sigma_s
    \end{pmatrix}={\rm Wi}\begin{pmatrix}
        G_\parallel&G_\perp\\
        -G_\perp&G_\parallel
    \end{pmatrix}\begin{pmatrix}
        e_n\\
        e_s
    \end{pmatrix}\label{eq:matrixfourier
    sigma}
\end{equation}
yields
\begin{subequations}
\begin{align}
    G_\parallel=\,&\frac{({\rm De}-i)}{({\rm De}-i)^2-4\omega_B^2}\,{\rm De} \\  
    G_\perp= -&\frac{2i\omega_B}{({\rm De}-i)^2-4\omega_B^2}\,{\rm De},
\end{align}\label{eq:complexVisco}%
\end{subequations}%
where we introduce the Deborah number ${\rm De}=\omega \tau$ for frequency scaling.
Thus, the frequency response of the viscoelastic odd fluid is richer than the usual Maxwell model with the interplay of the two poles. Namely, in the shear sector, the elastic response is in opposition of phase, \emph{i.e.} $G'_\parallel<0$, up until  ${\rm De}_{I}=\scriptstyle\sqrt{4\omega_B^2-1}$ (thus possible for $\omega_B\geq1/2$), where it becomes a purely viscous fluid. In a similar way, the odd sector has a pure elastic behaviour at the ${\rm De}_{II}=\scriptstyle\sqrt{4\omega_B^2+1}$ resonance
. These transitions are of paramount importance for the oscillatory analysis we bring forth, and are a clear manifestation of odd viscosity on a given fluid. %
{The implications of this coupling are illustrated in Fig. \ref{fig:intro}C, where one can observe the change in moduli of both sectors for increasing $\omega_B$. Where it is evident that the rotational consequences are not restricted to the odd response, but extend to the conventional shear rheology itself. Note, in particular, that in the shear sector there are regions where $G'$ drops but $G''$ increases, a fact that we will resort to later while discussing fluidisation.} In the coming section, we shall analyse how to exploit such resonant and nonlinear attributes of the model to unlock further rheological properties.

\section{Towards Controlled Odd Rheology}

The nondissipative nature of odd viscosity, albeit of great interest, also makes it harder to probe experimentally. It is well known that in incompressible flows the odd viscous force can be cast as the gradient of vorticity and thus absorbed into a corrected pressure, so that bulk steady-state experiments have little sensitivity to odd phenomena; odd stresses then become observable only through boundary conditions, most notably at free surfaces~\cite{Soni2019TheFluid,Franca2026,duPlessis2026}. The frequency response, on the other hand, is considerably rich, and the natural experimental setting to investigate odd viscoelastic materials is to impose an oscillatory shear~\cite{Hyun2011ALAOS,Giacomin1993,Ewoldt2008NewShear} such that 
\begin{equation}
    \dot e_n=0,\quad \dot e_s= \cos\big( {\rm De}\,  t\big).
\end{equation}
This setting exposes an interesting fingerprint: with $e_n=0$, $\sigma_n = G_\perp e_s$, so the odd fluid develops a normal-stress difference at the fundamental frequency, linear in the strain amplitude and governed by the moduli $G'_\perp$, $G''_\perp$ (Fig.~\ref{fig:intro}C). 
Note that in conventional viscoelastic fluids the normal-stress difference is even under strain reversal and hence shows up only at second order, as a steady offset plus a second harmonic~\cite{Bird1987}.
A first-harmonic normal stress here is therefore a direct (potentially experimental) rheometer-level signature of odd viscoelasticity.
Evidently, the forcing deformation rate can be imposed on the normal component instead of the shear one, but in experimental settings, it is less natural to do so.


Since the parity breaking underlying an odd response must be maintained by an external field or drive, or by internal activity, it is natural to consider that the circulation that imposes the parity breaking can evolve in time with a period $T$ such that $\omega_B(t+T)=\omega_B(t)$. With this condition, the monodromy matrix~\cite{V.A.Yakubovich1975Linear1} takes the form 
\begin{equation}
    \mat{X}(T)=e^{-T}\begin{pmatrix}
        \cos a& \sin a\\
        -\sin a& \cos a
    \end{pmatrix},
\end{equation}
with $a=\int_0^T 2\omega_B(s)\,{\rm d}s$, the accumulated coupling over a period. It is then clear that the Floquet multipliers necessarily lie inside the unit circle and that the system response remains bounded even for the inhomogeneous case~\cite{V.A.Yakubovich1975Linear1} with an applied strain rate. Such periodic evolution of $\omega_B$ can be thought to derive from fully internal processes of the fluid particles, in a characteristic way of active matter, but it is still plausible to set a desired modulation frequency through careful design of a synthetic material. Another possibility is to rely on driving the system as desired, not so differently from what is already done to attain odd fluid ~\cite{Soni2019TheFluid}. Such an approach could then unlock true parametric control of the rheology in both sectors.
This is, in effect, a classical and dissipative instance of Floquet engineering~\cite{Oka2019Floquet}, where the periodic drive does not only perturb the response but generates an effective rheology, with a sideband structure derived below, that is absent in the static fluid.
Note, although global stability is guaranteed, the evolution of the stress tensor under oscillatory coefficients is far from simple and can display interesting behaviour as the orbit in the stress-strain-strain rate space~\cite{Rogers2019,Ewoldt2008NewShear,Giacomin1993} gets deformed by responding to beating frequencies. The undulation imposed by the said beating frequency when projected on the strain--stress plane leads to strain softening/stiffening and thickening/thinning. That is, under the combined effect of the parametric oscillation of $\omega_B$ and the driving imposed at $\dot e_s$, the trajectory, $\vec{s}(t)=( e_s(t),\dot e_s(t),\sigma(t))$, of the system state on the space of strain, strain rate, and stress can no longer be contained in a plane. Resorting to the formalism of the sequence of physical processes (SPP)~\cite{Rogers2019,Rogers2017InViscoelasticity,Choi2019TheShearing}, the instantaneous dynamic moduli,  
\begin{equation}
    G_t'=\pderiv{\sigma}{e}=\frac{\dot \sigma \,{\rm De}\cos( {\rm De}\,t)-\ddot \sigma \sin( {\rm De}\,t)}{{\rm De}} \quad\text{and}\quad G_t''= \omega\pderiv{\sigma}{\dot e}=-\frac{\dot \sigma \,{\rm De}\sin( {\rm De}\,t)+\ddot \sigma \cos( {\rm De}\,t)}{{\rm De}}  , \label{eq:SPPmoduli}
\end{equation}
can be defined from the binormal vector of said trajectory ~\cite{Rogers2017InViscoelasticity}, one arrives at a scenario where the instantaneous elasticity and viscosity evolve in time in a convoluted fashion. On the parametric control scenario, there is a vast freedom in the choice of $\omega_B(t)$ since stability is guaranteed, and the discussion can be framed into the search of an optimal control to satisfy a given task or desired response. As a proof of concept, we consider a simple raised-cosine $\omega_B(t)=\omega_B\cos^2 2\pi t/T$ excitation. The rationale behind this form of $\omega_B(t)$ is to emulate a situation where microscopic elements of the fluid cyclically transition from a state of no rotation to another of maximal rotation. 

\begin{figure}[!ht]
    \centering
\includegraphics[width=\linewidth]{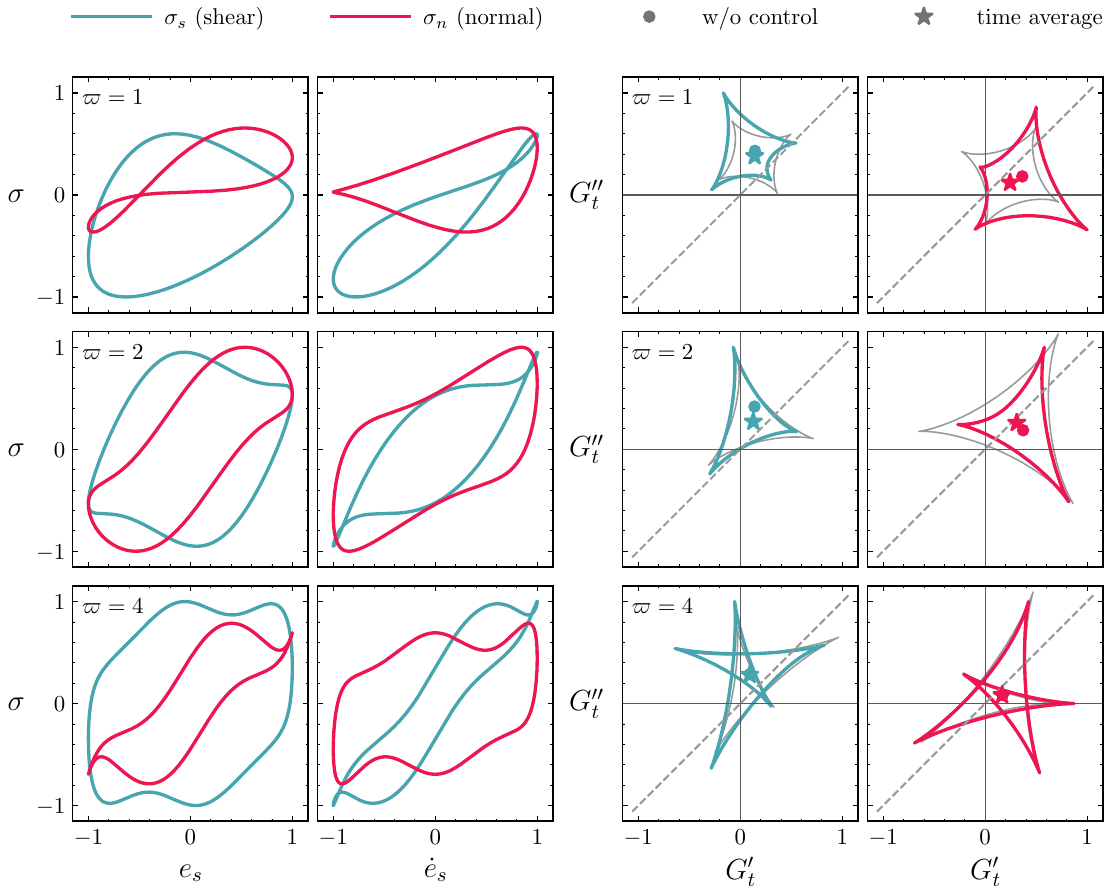}
    \caption{Lissajous curves and Cole--Cole diagrams for the system under the parametric modulation $\omega_B(t)=\omega_B\cos^2(2\pi t/T)$ for $\varpi=4\pi/T=1$ (top row), $2$ (middle row) and $4$ (bottom row). All numerical solutions use $\mathrm{De}=1$,  $\omega_B=1$ and $\mathrm{Wi}=1$. Shear (teal) and normal (magenta) stress response to oscillatory shear strain $e_s$ (first column) and strain rate $\dot{e}_s$ (second column). Cole--Cole diagrams of the instantaneous moduli, dashed line indicating the yield/reformation line of $G'=G''$, i.e. above such line the material is viscous fluid-like and below elastic solid-like. The modulus for constant coupling at its cycle mean, $\omega_B/2$, given by \eqref{eq:complexVisco}, is marked by a circle (w/o control), while the time average of $G'$ and $G''$ is marked by a star; both markers follow the colour of their stress sector. Thin grey curves in both Cole--Cole sectors show the approximation to the complex stress up to the first sideband, using the same scale as the numerical curves; their shear-sector moduli are given by ~\eqref{eq:shearUpperSideband} and~\eqref{eq:shearLowerSideband}. The storage and loss axes within each Cole--Cole panel use the same normalisation.}
        \label{fig:lissajousParametric}
\end{figure}

{
The implications of any parametric control of the rotational coupling are better understood as a phase modulation of $\vec{\sigma}$; to illustrate that, let us cast the stress and strain rate components as complex scalars $\Sigma\triangleq\sigma_n+i\,\sigma_s,\text{and}\, \dot E\triangleq \dot e_n+i\,\dot e_s$. Equation \eqref{eq:coupledMaxwell} then reads 
\begin{equation}
\big[\partial_t+1+2i\omega_B(t)\big]\Sigma={\rm Wi}\,\dot E
\end{equation}
and its solution after the transients subside is given by $\Sigma(t)={\rm Wi}\int_{-\infty}^t  \mathcal{G}(t,t') \dot E(t')\,{\rm d}t'$ where we introduced the retarded Green function
\begin{equation}
    \mathcal{G}(t,t')=\Theta(t-t')\exp\big[-(t-t')\big]\exp\left[-i\int_{t'}^t 2\omega_B(s)\, {\rm d}s\right], 
\end{equation}
in which the first exponential term provides the expected viscoelastic memory decay and the second one accounts for the rotation of the stress during the memory time; once again showing that the system is always stable under parametric control. 
Plugging in our raised-cosine ansatz 
\begin{equation}
    \mathcal{G}(t,t')=\Theta(t-t')e^{-(t-t')}e^{-i(t-t')\,\omega_B}e^{-iw\sin\varpi t}e^{iw\sin\varpi t'}
\end{equation}
with $\varpi=4\pi/T$ the parametric excitation frequency and $w=\omega_B/\varpi$. Recalling that we are interested in a harmonic strain rate {, $\dot E=i\cos(\mathrm{De}\,t)
=i(e^{-i\mathrm{De}t}+e^{i\mathrm{De}t})/2$, }and resorting to the Jacobi--Anger expansion of the trigonometric exponentials, we then arrive at 
{, 
\begin{equation}
    \Sigma(t) = \frac{i\,\mathrm{Wi}}{2}
    \sum_{\ell=-\infty}^{\infty}
    \Big[
        C_\ell(\mathrm{De})e^{-i(\mathrm{De}+\ell\varpi)t}
        +C_\ell(-\mathrm{De})e^{-i(-\mathrm{De}+\ell\varpi)t}
    \Big],\,\text{with}\,\,\,
    C_\ell(\xi) =\!\!\!
    \sum_{n=-\infty}^{\infty}\!\!
    \frac{J_{n+\ell}(w)J_n(w)}
    {1-i(\xi-\omega_B-n\varpi)},
\end{equation}%
}%
%
where $J_n$ is the Bessel function of the first kind and order $n$. This explicitly shows that the parametric oscillation of frequency $\varpi$ creates sidebands around the driving central frequency of ${\rm De}$ as a frequency comb with the amplitude governed by the Bessel coefficients. %
If we look for the first sidebands contributions, that is $\ell\in\{0,+1,-1\}$, splitting $\Sigma\simeq\Sigma_0+\Sigma_++\Sigma_-$, with $\Sigma_\pm$ denoting the components at frequencies ${\rm De}\pm\varpi$, we can compute the SPP moduli. While $\Sigma_0$ results on a single point of $G_t$, as the time dependence drops out, both the side bands retain the time dependence and trace a trajectory on the complex $G$ plane. 
{%
For the upper sideband $\Sigma_+$,~\eqref{eq:SPPmoduli} then yields, for the shear sector, 
\begin{equation}
    G'_t+iG''_t
    =\frac{\mathrm{De}+\varpi}{2\mathrm{De}}
    \Big[
        \varpi B_{+1}
        e^{-i(2\mathrm{De}+\varpi)t}
        {}-{}(2\mathrm{De}+\varpi)
        B_{+1}^{*}e^{i\varpi t}
    \Big],
    \label{eq:shearUpperSideband}
\end{equation}
while for $\Sigma_-$
\begin{equation}
    G'_t+iG''_t
    =\frac{\varpi-\mathrm{De}}{2\mathrm{De}}
    \Big[
        (2\mathrm{De}-\varpi)
        B_{-1}^{*}e^{-i\varpi t}\\
        {}+{}\varpi B_{-1}
        e^{-i(2\mathrm{De}-\varpi)t}
    \Big],
    \label{eq:shearLowerSideband}
\end{equation}
where the two frequency branches are incorporated into
$B_\ell=i\,\mathrm{Wi}
    \left[
        C_\ell(\mathrm{De})+
        C_{-\ell}^{*}(-\mathrm{De})
    \right]/2.$
The analogous normal-sector response follows by taking the real part of the same truncated $\Sigma$.
These expressions are no more than the sum of two circular motions
.
}

Thus, the instantaneous moduli describe hypocycloids (cf. Fig.~\ref{fig:lissajousParametric}) in the complex plane\footnote{Commonly named Cole--Cole diagrams, it is important to note that, contrary to the more well-known dielectric Cole--Cole plots, here the parametrisation occurs over time and not frequency.}, when Deborah number and parametric and half Floquet frequency $2\pi/T$ are commensurable, producing curves with $p$ number of cusps, where $p$ is determined by
\begin{equation}
    \frac{{\rm De}\,T}{2\pi}+2=\frac{p}{q},
\end{equation}
with irreducible $p/q$.
{
The complete response, however, is not restricted to the first
sideband, and the superposition of the remaining harmonics
deforms the ideal hypocycloidal trajectories (cf. Fig.~\ref{fig:lissajousParametric}).
Nevertheless, the characteristic cusp counting is preserved
over the parameter range explored here.
}
The analysis of the hypocycloidal trajectories shows that the parametric oscillation itself is capable of rendering the material an elastic solid, with $G_t'> G_t'' $, in parts of the trajectory, especially for higher modulating frequency. Also, the average dynamic moduli are displaced compared to the theory of constant rotation given by \eqref{eq:complexVisco}. This constitutes further proof of the complex behaviour of the fluid. 
}

Until now we discussed how the microscopic parity breaking unfolds into a viscoelastic fluid where shear and odd sectors are intertwined. And we saw the linear and parametric response of such a system. But, an underlying assumption was that the microscopic rotation or transverse forces were always present and independent of the mesoscopic state of the fluid. However, it is not unrealistic to think of a circumstance where the rate of rotation depends on the applied stress or deformation rate. 
For instance, it is reasonable to consider that such symmetry breaking is frozen below a critical stress level. That is, the microscopic rotation only occurs after the particles are sufficiently strained, which translates to a Bingham-like model of the form 
{
\begin{equation}
    \omega_B(\vec{\sigma})=\begin{cases}
        0&, |\vec{\sigma}|<\sigma_0\\ 
        \omega_B &, |\vec{\sigma}|\geq\sigma_0
    \end{cases}\doteq\omega_B\,\Theta\left(\sqrt{\sigma_s^ 2+\sigma_n^2}-\sigma_0\right),\label{eq:binghm}%
\end{equation}%
}%
where we used the determinant of $\vec \sigma$ as its second invariant in two dimensions. 
Evidently, such a transition challenges the conservation of angular momentum in the fluid, and it has further implications for the full dynamics of the system at the level of the transport equations. However, transients of angular momentum transport are usually short-lived~\cite{Kagan1962ONFREEDOM,Curtiss1956KineticMolecules}. Also, in the present work, we focus on the rheological properties and leave aside the study of the implications at the Navier--Stokes level for future research.

With this model, the shear and normal stresses only couple when the stress magnitude surpasses the critical yield stress level $\sigma_0$, which we here take to be the same as the stress normalisation. For this reason, we can expect the stress signal to be well represented by its first Fourier harmonics, writing
\begin{equation}
    \sigma(t)=\frac{\rm Wi}{\rm De}\sum_{\text{odd}\, m}G'_m\sin m\,{\rm De}\, t+ G''_m\cos m\,{\rm De}\, t = \sum_{\text{odd}\, m} I_m\sin\big(m\,{\rm De}\, t+\delta_m \big),
\end{equation}
for each sector, with the first harmonic $G_1\approx G(\omega)$ from the linear theory, and higher-order contributions being a consequence of the step transition so much that the ratio $|I_3/I_1|$ is a proxy for how far from linearity the system is behaving.

\begin{figure}[!ht]
    \centering

\includegraphics[width=\linewidth]{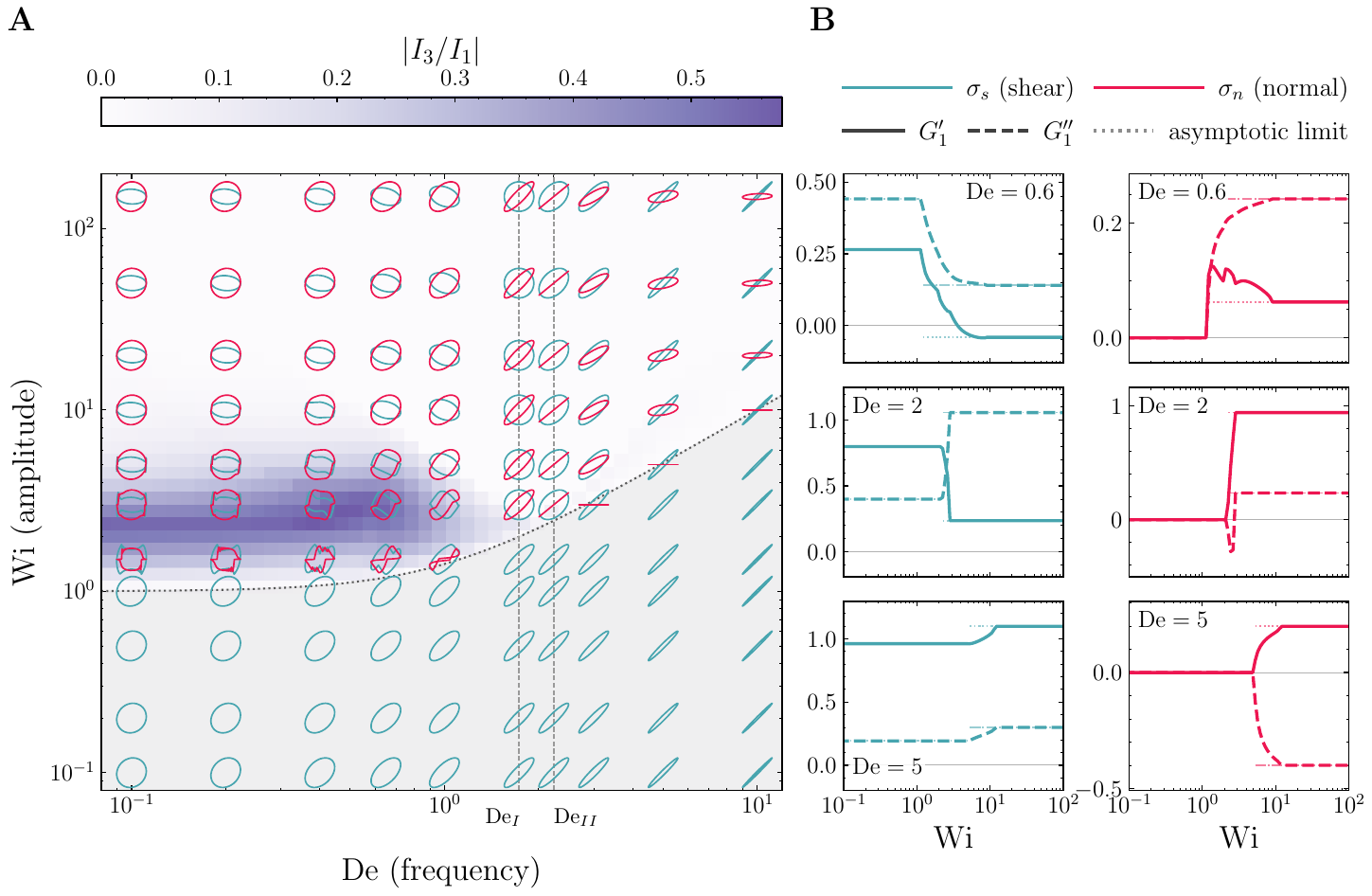}

    \caption{\textbf{A.} Odd Pipkin diagram for the rotational yielding model. The Lissajous curves are normalised to their extrema and placed centred over the corresponding Deborah and Weissenberg parameters. The teal curves represent the shear sector $\sigma_s$ and the magenta the normal $\sigma_n$. The background cells show the shear-sector ratio $|I_3/I_1|$ at the sampled $(\mathrm{De},\mathrm{Wi})$ points, without interpolation. The rotational step is numerically regularised as $[1+\tanh((|\vec{\sigma}|-\sigma_0)/\epsilon)]/2$ with $\epsilon=10^{-4}$ in normalised stress units. \textbf{B.} First-harmonic moduli in the shear (left) and normal (right) sectors at $\mathrm{De}=0.6$, $2$ and $5$ from top to bottom; solid and dashed curves show $G_1'$ and $G_1''$, respectively, while the light reference curves indicate the asymptotic limits.
    {For the representative choice $\omega_B=1$, the upper boundary of fluidisation is ${\rm De}_c\simeq2.751$.}} 
    \label{fig:pipkin}
\end{figure}

As a consequence of the rotational yield, for low-amplitude strains (i.e. for ${\rm Wi}<\sqrt{1+{\rm De}^2}$, as clearly confirmed by Fig.~\ref{fig:pipkin}) the material loses its odd properties, and one gets $\sigma_n=0$ and a regular Maxwell material regarding the shear stress. In the opposite regime, the large amplitude of the driving ensures that $\omega_B\neq0$ throughout most of the cycle and the moduli asymptotically approach their values of the odd linear scenario  (Fig. \ref{fig:pipkin}). 
%
{Remarkably, the activation of the coupling through the critical stress can also reverse the state of fluidisation. Such a reversal can be compactly identified by the condition
$\left.(G''_\parallel-G'_\parallel)\right|_{\omega_B=0}
\left.(G''_\parallel-G'_\parallel)\right|_{\omega_B\neq0}<0$, which for the coupled Maxwell model herein discussed reduces to the region $1<{\rm De}<{\rm De}_c$ with the critical Deborah defined as the real root of the cubic 
\begin{equation}
    {\rm De}_c^3-{\rm De}_c^2+(1-4\omega_B^2){\rm De}_c-(1+4\omega_B^2) =0.
\end{equation}

For a given value of $\omega_B$ this defines a finite rotation-induced fluidisation window. Indeed, in the absence of odd coupling, the usual Maxwell response for ${\rm De}>1$ has $G'_\parallel>G''_\parallel$, corresponding to an elastic-dominated, solid-like response. However, activation of the rotational degree of freedom by the crossing of the Bingham condition of \eqref{eq:binghm} reverses this ordering, such that unlocking the microscopic rotation dynamically fluidises an otherwise solid-like response. Further down, for ${\rm De}>{\rm De}_c$, the odd coupling is no longer sufficient to reverse the ordering of the two moduli and the response remains elastic-dominated.
Thus the two boundaries of this window have distinct origins. At ${\rm De}=1$, the uncoupled Maxwell fluid already lies at the crossover $G'_\parallel=G''_\parallel$, so that any finite rotational coupling shifts the response towards the loss-dominated regime. Conversely, at ${\rm De}={\rm De}_c$, the equality $G'_\parallel=G''_\parallel$ is reached on the rotationally activated branch. Thus, increasing ${\rm De}$ across the fluidisation window continuously transfers the crossover from the locked to the unlocked state. 

Beyond providing a further signature of the odd coupling, this window suggests a direct route towards controllable rheology. In the present model, the rotational degree of freedom is unlocked only after the local stress exceeds a prescribed threshold, so that the material itself provides an intrinsic feedback mechanism: increasing the forcing can activate microscopic rotation and thereby drive the macroscopic response from elastic-dominated to viscous-dominated. In this sense, the internal chiral dynamics acts as a state-dependent rheological control parameter rather than a passive material property.
The same principle also admits an externally programmable realisation. If the microscopic rotation can be controlled independently, $\omega_B$ provides a direct handle with which to move the material across the boundary $G'_\parallel=G''_\parallel$ at fixed driving frequency. Moreover, promoting this control parameter to a time-dependent quantity, $\omega_B(t)$, extends the static fluidisation mechanism into the parametric regime discussed above. 

}

%

Additionally, solving \eqref{eq:coupledMaxwell} with \eqref{eq:binghm} crossing the yield criterion further unlocks the nonlinear nature of the system, rendering it admittedly distinct from the linear case, with deformed curves in the stress--strain plane. The amplitude ratio between the third and first harmonic modes of the stress on the Pipkin diagram of Fig.~\ref{fig:pipkin} shows the said nonlinear stress features peak around ${\rm Wi}\sim 3$, ${\rm De}\lesssim 0.6$ (for $\omega_B=1$) from a balance of sufficiently large amplitude for a significant part of the orbit to surpass the yield at $\sigma_0$ but not excessively or it would fall into the linear limit again. Likewise, the driving frequency needs to be fast enough for the stiffening  cycles to be swift and not quasi-static but still below the resonances of ${\rm De}_{I}$ and ${\rm De}_{II}$. 
There are two reasons for this. First, the relative intermediate frequency prevents a separation of time scales, and the system thus becomes more responsive to the time history of the applied shears; second, in the odd sector, the viscous and elastic responses are of comparable strength and in phase opposition to the shear sector, facilitating the evolution of spectrally richer responses, precursors of nonlinear behaviour.

\section{Conclusion \& Outlook}

Research on odd viscosity and odd elasticity have progressed largely in parallel, with only a few works addressing their viscoelastic interplay, and even fewer exploring the possibility of transient fluidisation. However, with recent and rapidly advancing experimental realisations of odd materials, it has become pressing to consider such effects from a theoretical perspective and to explore the rich and diverse rheology they unlock. 
Our work provides a basis for considering the control of the rheological state of a chiral fluid via the microscopic rotations of its components and therefore coupling the odd and shear sectors of the dynamic modulus.

We show that relatively simple manipulations of the shear-normal coupling deeply impact the overall response of the fluid. In particular, the naturally emerging coupled Maxwell model, although stable under parametric excitations, drastically changes its instantaneous response moduli as discussed. Thus, it provides us with a way to control the properties of the fluid by carefully choosing the time evolution of the rotational coupling $\omega_B$. As proof of concept, we chose a raised sinusoidal switch function, which already shows the overall behaviour of cyclic shear thinning/thickening. Yet, if a certain rheological response is desired, more complex time evolutions can be determined.
Note, nonetheless, that such parametric variation need not be externally imposed but could instead be encoded in the material itself (from micro or mesoscopic engineering), thus achieving a functional material capable of adaptability or driven by internal synchronisation. 
Similarly, the reported transient fluidisation from the rotational yielding can be key to future materials exhibiting complex rheological responses from a tailored rotational transition model. Since, the coupling may not only be a periodic function of time but display instead an intrinsic response to the stress of the material. Here we showed that a Bingham-like model for the coupling unlocks as well periodic cycles of  softening--thinning and stiffening--thickening that can be potentially tailored to respond to the environment.

The stress-dependent unlocking considered here and the periodic modulation of the odd coupling are therefore two limits of the same control strategy. The former provides an intrinsic, stimulus-responsive modulation of the rheology; such a mechanism is reminiscent of the broader logic of adaptive or smart materials, in which an internal degree of freedom reorganises in response to an applied stimulus and changes the effective mechanical response. The latter, by contrast, allows the mechanical response to be actively programmed in time. Their combination suggests the possibility of odd viscoelastic materials whose effective solid- or fluid-like character can be selected, switched, or periodically modulated through control of their internal rotational dynamics. 
%
%
Both limits of this control strategy have natural experimental counterparts, one driven and one active. In magnetised two-dimensional electron fluids, $\omega_B=\omega_c\tau$ is the Hall parameter set by the field, the precession of the stress at $2\omega_c$ underlies the shear Bernstein modes~\cite{bernsteinOdd}, and a time-dependent field realises directly the parametric modulation (Sec.~III). Among active systems, the recently introduced suspensions of spinning microalgae~\cite{eisenmann2026chirality} are particularly well suited, where cells bind into transient clusters that orbit collectively, so that the stress carried by the network rotates by an angle of order unity before the bonds that carry it break; this leads to the regime of strong odd coupling considered here, consistent with the observation that their measured odd diffusivity is comparable to or larger than its even counterpart~\cite{eisenmann2026chirality}, although a complete connection to odd stress coupling remains to be established. 

The natural next step is to incorporate such models into the full Cauchy momentum equation of the fluid, where the full potential of the complex rheology that we bring forward can unfold. It will be of particular interest to observe how secondary nonlinearities may appear from the objective time derivative of the stress. 
With that, normal and shear stresses are further coupled, but via second-order rheology, contrary to the effects that we discuss herein, which arise from a first-order viscoelasticity model. More broadly, these results point towards metafluids whose rheology is controlled through the rotational state of the constituents, with imposed, modulated, or stress-activated rotational dynamics as the design variable.


\section*{Acknowledgements}

We thank Corentin Coulais for fruitful discussions. We acknowledge support from the ERC Grant No. “2023-StG-101117025, FluMAB.” This publication is part of the Vidi project Living Levers with File No. 21239, financed by the Dutch Research Council (NWO). 

\section*{Author contributions}
P.C.: Conceptualization, Methodology, Investigation, Formal analysis, Writing - original draft - review \& editing
M.J.: Conceptualization, Supervision, Funding acquisition, Writing – review \& editing.

\section*{Use of generative AI}
Claude (Anthropic; Claude Opus 5.5 and Claude Fable 5.1) was used to improve grammar and readability, to draft analysis code, and to numerically cross-check the analytical expressions. All AI-assisted content was reviewed, verified and revised by the authors, who take full responsibility for the manuscript.

\printbibliography

\pagebreak
\appendix
\renewcommand{\thesection}{\Alph{section}}
\titleformat{\section}
  {\bf\sffamily}
  {APPENDIX \thesection:}
  {5pt}
  {\MakeUppercase}

\clearpage
\section{Objective transport of the stress tensor}
\label{APP:objective}
{
In what follows, we set the mass of the particles as $m\equiv 1$.

We want to retrieve the evolution of the stress-pressure tensor $\Pi$ defined as the covariance of the velocities, i.e. 
\begin{equation}
    \Pi_{ij}=\langle c_ic_j \rangle =\int c_i c_j f\,{\rm d}\vec{v}
\end{equation}
where $\vec{c}$ is the peculiar velocity $\vec{c}=\vec{v}-\langle\vec{v}\rangle=\vec{v}-\vec{u}$, and with the stress-pressure tensor having the decomposition $\Pi_{ij}=p\delta_{ij}-\sigma_{ij}$ into its hydrostatic pressure and stress parts.
From the Boltzmann equation 
\begin{equation}
\ptderiv{f}+\vec{v}\cdot\bm{\nabla}_\vec{x}f+\vec{a}\cdot\bm{\nabla}_\vec{v}f=\mathfrak{S},
\end{equation}
in which we assume that the acceleration $\vec{a}$ includes both physical and fictitious contributions; multiplying by $c_ic_j$ and integrating over the microscopic velocities leads to the transport law
\begin{equation}
\ptderiv{\Pi_{ij}}+\partial_k(\Pi_{ij}u_k)+\partial_k\langle c_ic_jc_k\rangle + \Pi_{jk}\partial_ku_i+ \Pi_{ik}\partial_ku_j -\langle c_ia_j+c_ja_i\rangle-\langle c_ic_j \partial_{v_k} a_k\rangle=\langle c_ic_j\mathfrak{S}\rangle.\label{APP:eq:evolvePi}
\end{equation}
From the Grad moments expansion, we obtain the closure $\langle c_ic_jc_k\rangle=0$, and it is now also clear that if the acceleration is independent of the microscopic velocity, then the averages of the last two terms on the LHS vanish. However, if we allow for some linear contribution of $\vec{c}$ we can cast 
\begin{equation}
    \langle c_ia_j+c_ja_i\rangle+\langle c_ic_j \partial_{v_k} a_k\rangle\equiv\Omega_{ijkl}\sigma_{kl}.
\end{equation}
Further, under BGK approximation $\mathfrak{S}=(f^{(0)}-f)/\tau$ it follows immediately that 
\begin{equation}
    \langle c_ic_j\mathfrak{S}\rangle  =  \frac{\sigma_{ij}}{\tau};
\end{equation}
plugging the pressure and stress decomposition into \eqref{APP:eq:evolvePi} with these previous definitions results in 
\begin{equation}
\frac{D\sigma_{ij}}{Dt}+ \sigma_{ij}\partial_ku_k+ \sigma_{ik}\partial_ku_j+ \sigma_{jk}\partial_ku_i-\sigma_{kl}\partial_lu_k\delta_{ij} +\Omega_{ijkl}\sigma_{kl}+\frac{\sigma_{ij}}{\tau}=p\big(\partial_iu_j+\partial_ju_i-  \delta_{ij}\partial_ku_k  \big).
\label{APP:eq:evolveSigma}
\end{equation}
As mentioned, only forces explicitly depending on velocity contribute to the averaged terms; thus, to take into account the non-inertial nature of the transport, it is only necessary to consider the Coriolis contribution arising from a rotating frame
\begin{equation}
    a^\text{Cor}_i=-2W^o_{ij}v_j\quad\text{with}\quad \vec{W^o}=\vec{R}^\mathsf{T}\dot{\vec{R}},
\end{equation}
which then gives us
\begin{equation}
    \langle c_ia^\text{Cor}_j+c_ja^\text{Cor}_i\rangle+\langle c_ic_j \partial_{v_k} a_k^\text{Cor}\rangle = 2W^o_{ik}\sigma_{kj}-2\sigma_{ik}W^o_{kj}.
\end{equation}

\begin{subequations}
\begin{equation}
\overset{\circledcirc}{\sigma}_{ij} +\Omega_{ijkl}\sigma_{kl}+\frac{\sigma_{ij}}{\tau}=p\big(\partial_iu_j+\partial_ju_i-  \delta_{ij}\partial_ku_k  \big),
\end{equation}
with 
\begin{equation}
   \overset{\circledcirc}{\sigma}_{ij}  \equiv \frac{D\sigma_{ij}}{Dt}+ \sigma_{ij}\partial_ku_k+ \sigma_{ik}\partial_ku_j+ \sigma_{jk}\partial_ku_i-\sigma_{kl}\partial_lu_k\delta_{ij} + 2W^o_{ik}\sigma_{kj}-2\sigma_{ik}W^o_{kj}.\label{eq:ObjectiveDeriv}
\end{equation}
\end{subequations}
Introducing, in this way, an unconventional differential operator for the transport of the stress tensor. To confirm that such an operator is in fact objective,
consider two observers related by an arbitrary time-dependent rigid-body transformation
\begin{equation}
x_i^*=Q_{ia}(t)x_a+b_i(t),    
\end{equation}
where $\vec{Q}\in SO(2)$ is a proper orthogonal tensor, $
Q_{ia}Q_{ja}=\delta_{ij}.$
While the stress transforms as a tensor $\sigma_{ij}^*
=
Q_{ia}Q_{jb}\sigma_{ab}$, the velocity measured by the transformed observer follows by differentiating the transformation along a material trajectory and therefore,
\begin{equation}
u_i^*
=
Q_{ia}u_a+\dot Q_{ia}x_a+\dot b_i;
\end{equation}
and since the spatial derivative transforms as $
\partial_i^*=Q_{im}\partial_m$ the velocity gradient gives
\begin{equation}
\partial_i^*u_j^*
=
Q_{im}Q_{jn}\partial_m u_n
+
Q_{im}\dot Q_{jm}=
Q_{im}Q_{jn}\partial_m u_n-A_{ij},
\label{eq:grad_u_step}
\end{equation}
where we found it convenient to introduce the antisymmetric tensor 
$
A_{ij}
\equiv
\dot Q_{im}Q_{jm}=-A_{ji},    
$
so that it represents the angular velocity associated with the change of observer. Given the orthogonality of $\vec{Q}$ and antisymmetry of $\vec{A}$, it also follows trivially that 
\begin{equation}
   \partial_i^*u_i^*
=\partial_iu_i.
\end{equation}
Moreover, recall that the rotation rate $W^o_{ij}$ transforms as a rotational connection; thus,
\begin{equation}
W_{ij}^{o*}
=
Q_{im}Q_{jn}W^o_{mn}
-A_{ij}.
\label{eq:Wo_transformation}
\end{equation}

We now consider the different terms in the stress derivative defined in \eqref{eq:ObjectiveDeriv}. 
The material derivative of the stress transforms as
\begin{equation}
\frac{D^*\sigma_{ij}^*}{Dt}
=
Q_{im}Q_{jn}\frac{D\sigma_{mn}}{Dt}
+
A_{ik}\sigma_{kj}^*
+
A_{jk}\sigma_{ik}^*,
\label{eq:material_transformation}
\end{equation}
while the velocity-gradient contributions give us 
\begin{align}
    \sigma_{ik}^*\partial_k^*u_j^*
=
Q_{im}Q_{jn}\sigma_{mp}\partial_pu_n
+
A_{jk}\sigma_{ik}^*.
\label{eq:first_gradient_transformation}\\
\sigma_{jk}^*\partial_k^*u_i^*
=
Q_{im}Q_{jn}\sigma_{np}\partial_pu_m
+
A_{ik}\sigma_{kj}^*.
\label{eq:second_gradient_transformation}
\end{align}
as such, the material derivative and the two velocity-gradient terms together generate the additional observer-dependent contribution
\begin{equation}
2A_{ik}\sigma_{kj}^*
+
2A_{jk}\sigma_{ik}^*.
\label{eq:extra_rotation}
\end{equation}
The term containing the velocity divergence is already objective because of the invariance of the trace, so
\begin{equation}
\sigma_{ij}^*\partial_k^*u_k^*
=
Q_{im}Q_{jn}
\left(
\sigma_{mn}\partial_pu_p
\right).
\label{eq:divergence_term_objective}
\end{equation}
Again from the symmetry of $\vec{\sigma}$ and antisymmetry of $\vec{A}$ the scalar contraction appearing in the isotropic term is objective 
\begin{equation}
\sigma_{kl}^*\partial_l^*u_k^*\delta_{ij}=Q_{im}Q_{jn}\sigma_{kl}\partial_lu_k\delta_{mn}.
\end{equation}

It remains to consider the terms containing $W^o_{ij}$. Using \eqref{eq:Wo_transformation},
\begin{equation}
2W_{ik}^{o*}\sigma_{kj}^*
=
2Q_{im}Q_{jn}W^o_{mp}\sigma_{pn}
-2A_{ik}\sigma_{kj}^*,
\label{eq:Wo_sigma}
\end{equation}
and likewise for the second term. Consequently,
\begin{equation}
2W_{ik}^{o*}\sigma_{kj}^*
-2\sigma_{ik}^*W_{kj}^{o*}
=
Q_{im}Q_{jn}
\left(
2W^o_{mp}\sigma_{pn}
-2\sigma_{mp}W^o_{pn}
\right)
-2A_{ik}\sigma_{kj}^*
-2A_{jk}\sigma_{ik}^*.
\label{eq:Wo_complete}
\end{equation}

The last two terms in \eqref{eq:Wo_complete} cancel exactly the observer-dependent terms in \eqref{eq:extra_rotation}. Therefore, all dependence on the relative rotation of the two observers disappears. %
With that, we  finally obtain
\begin{equation}
\overset{\circledcirc}{\sigma}_{ij}^{\,*}
=
Q_{im}Q_{jn}
\overset{\circledcirc}{\sigma}_{mn},
\label{eq:objectivity_result}
\end{equation}
hence the generalised derivative defined in \eqref{eq:ObjectiveDeriv} is objective.

}

\end{document}